\documentclass[12pt]{article}
\usepackage{times}
\usepackage{geometry}
\usepackage{graphicx}
\usepackage[utf8]{inputenc}
\usepackage{enumitem,amssymb}
\usepackage{ragged2e}
\newlist{thematic}{itemize}{8}
\setlist[thematic]{label=$\square$}
\usepackage{pifont}
\usepackage{wrapfig}

\begin{document}
\raggedright
\huge
Astro2020 Science White Paper \linebreak

Mapping out the time-evolution of exoplanet processes\linebreak

\normalsize
\noindent \textbf{Thematic Areas:} \hspace*{60pt} $\boxtimes$ Planetary Systems \hspace*{10pt} $\boxtimes$ Star and Planet Formation \hspace*{20pt}\linebreak
$\square$ Formation and Evolution of Compact Objects \hspace*{31pt} $\square$ Cosmology and Fundamental Physics \linebreak
  $\square$  Stars and Stellar Evolution \hspace*{1pt} $\square$ Resolved Stellar Populations and their Environments \hspace*{40pt} \linebreak
  $\square$    Galaxy Evolution   \hspace*{45pt} $\square$             Multi-Messenger Astronomy and Astrophysics \hspace*{65pt} \linebreak
  
\textbf{Principal Author:}

Name: Jessie Christiansen  
 \linebreak                                             
Institution: Caltech/IPAC-NExScI
 \linebreak
Email: jessie.christiansen@caltech.edu
 \linebreak
Phone: (626) 720-9649
 \linebreak
 
\textbf{Co-authors:} 
  \linebreak
  Charles Beichman, Caltech/IPAC-NExScI\\
  David R. Ciardi, Caltech/IPAC-NExScI\\
  Daniel Huber, University of Hawai`i \\
  Mark S. Marley, NASA ARC
  \linebreak

\textbf{Co-signers:} 
  \linebreak
  Chuanfei Dong, Princeton University \\ Eric Lopez, NASA GSFC\\ Courtney D. Dressing, University of California, Berkeley \\ Susan Mullally, STScI\\ Peter Plavchan, George Mason University  \\
 
Edwin Kite, University of Chicago  \\
  Jonathan Fortney, UCSC\\ Natalie Hinkel, Southwest Research Institute\\ Derek Buzasi, Florida Gulf Coast University\\ 
  Andrew W. Mann, University of North Carolina at Chapel Hill\\
  Benjamin T. Montet, University of Chicago\\
  Gijs D. Mulders, The University of Chicago \\
  William Danchi, NASA GSFC \\
  Elisabeth R. Newton, Dartmouth College \\
  Tae-Soo Pyo, Subaru Telescope \\
  Kevin Hardegree-Ullman, Caltech/IPAC-NExScI \\
  Carey Lisse, Johns Hopkins University Applied Physics Lab \\
  Seth Redfield, Wesleyan University \\

\vspace{0.5in}

\justifying  
\textbf{Abstract:}

There are many competing theories and models describing the formation, migration and evolution of exoplanet systems. As both the precision with which we can characterize exoplanets and their host stars, and the number of systems for which we can make such a characterization increase, we begin to see pathways forward for validating these theories. In this white paper we identify predicted, observable correlations that are accessible in the near future, particularly trends in exoplanet populations, radii, orbits and atmospheres with host star age. By compiling a statistically significant sample of well-characterized exoplanets with precisely measured ages, we should be able to begin identifying the dominant processes governing the time-evolution of exoplanet systems.

\pagebreak

\section{Introduction}
\vspace{-10pt}
As the exoplanet field becomes increasingly well established, we are rapidly transitioning from characterisation of individual exoplanet systems to characterisation of exoplanet populations. There are several broad questions that motivate such analysis: How do planetary systems form? What features of their natal environment and host star(s) govern their subsequent migration and evolution? On what timescales do migration and evolution occur? The National Academies of Science Exoplanet Survey Strategy identified as the first goal of exoplanet science in the coming decade {\it "to understand the formation and evolution of planetary systems as products of the process of star formation, and characterize and explain the diversity of planetary system architectures, planetary compositions, and planetary environments produced by these processes."}\newline

Ideally, we would observe a statistical sample of exoplanets from birth through death, but astronomical timescales limit us to observing exoplanets at a short snapshot in time. However, with a large enough sample of exoplanets orbiting a well-characterized set of stars that span a wide range of ages, we can reconstruct the time evolution of planetary systems. In this white paper, we seek to identify the feasible observations that could be made in the next decade that would enable improving our understanding of exoplanet formation and evolution.  In Section \ref{sec:evolution}, we summarize the predicted timescales that will differentiate between competing planet formation and migration models. In Section \ref{sec:radius}, we describe the time-evolution, both observed and predicted, of the radii of transiting planets, identifying several open questions. In Section \ref{sec:orbits}, we describe how mapping out the orbital evolution of giant planets can provide an alternate path for identifying the dominant migration processes. In Section \ref{sec:atmos}, we detail the expected atmospheric evolution and the predicted signatures of that evolution. Finally, in Section \ref{sec:stars}, we describe the potential for asteroseismology to provide the necessary sample of precise stellar ages.


\section{Planet formation and evolution}
\label{sec:evolution}
\vspace{-10pt}
The population of known transiting planets is thus far consistent with the predicted outcome of the core accretion model of planet formation (Pollack et al., 1996; Ida \& Lin, 2004). In this model, gas-giants form in a multi-stage process, where a $\sim$10 M$_{\oplus}$ core forms from the coagulation of planetesimals, then rapidly accretes gas from the protoplanetary disk. This process must occur rapidly, on the timescale of the observed dissipation of the primordial disk $\lesssim$10 Myr (Haisch et al., 2001). For smaller rocky planets, this timescale is less severe. These planets form as an initial population of Moon- to Mars-mass bodies undergoes a phase of relatively slow chaotic growth via collisions over $\sim$50--100 Myr (Goldreich et al., 2004).\newline

Whether these planets formed {\it in situ}, very close to the star (Chiang \& Laughlin, 2013), or at wider orbital separations and then migrated inward (Schlaufman et al., 2009) can be inferred by detecting planets orbiting young stars of known age. Additionally, the orbital separations and ages of such planets can place constraints on the type of migration that may be responsible for their current separations (e.g. disk migration or scattering) and the timescale of that mechanism (Ida \& Lin, 2008; Ford, 2006). Disk-driven planet migration must occur early in the evolution of the star planet system while there is still substantial gas in the disk ($\lesssim$10 Myr), whereas migration due to scattering from a binary star companion or additional planets is much slower, and occurs over an extended period of time ($\sim$100 Myr--1 Gyr).\newline

\begin{wrapfigure}{r}{0.55\linewidth}
\Centering
\includegraphics[width=0.55\textwidth]{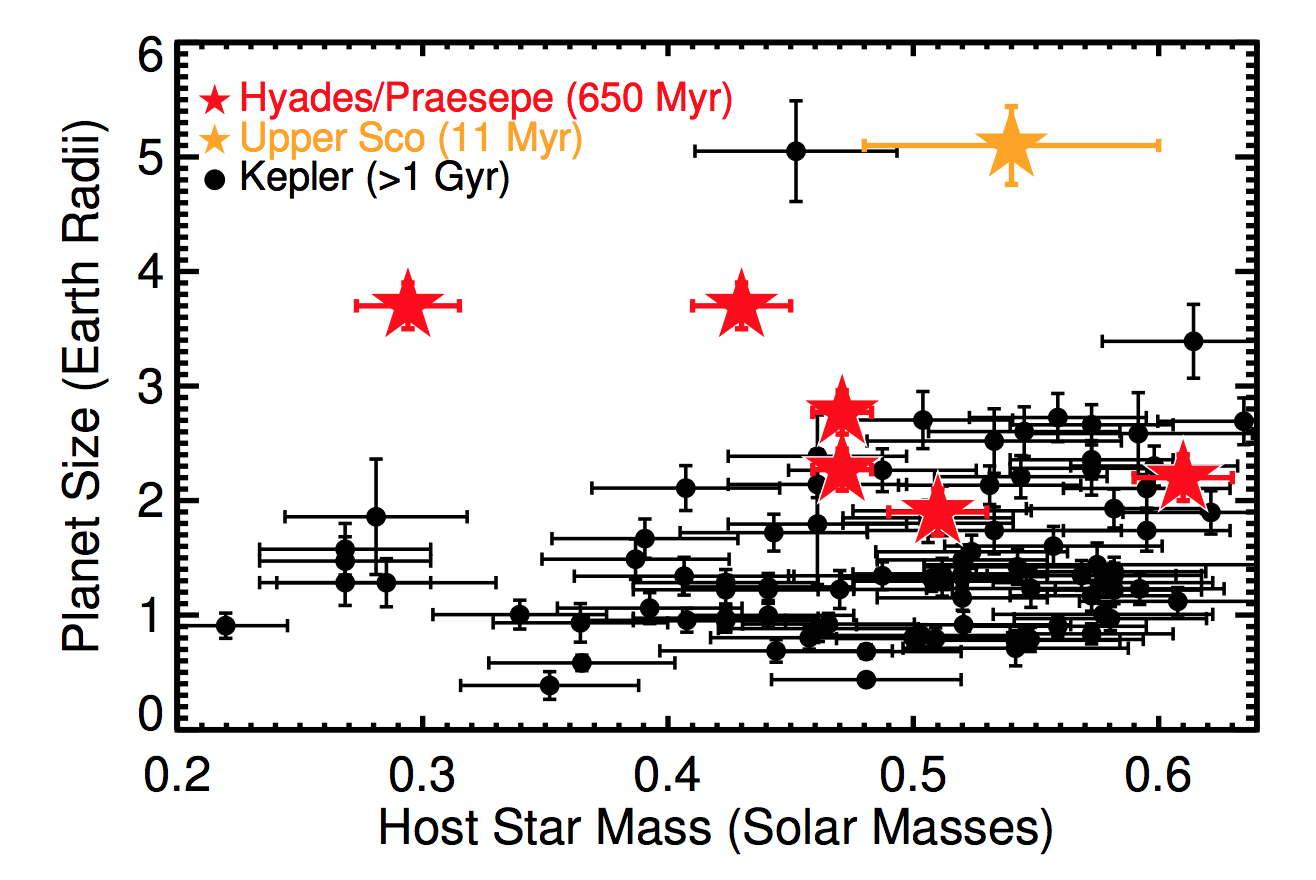}
\caption{Reproduced from Fig. 12 of Rizzuto et al. (2018), showing the radii of the planets detected around young stars compared to the {\it Kepler} field star sample.}
\label{fig:rizzuto_k2}
\end{wrapfigure}

The NASA {\it K2} mission has recently allowed us to explore the properties of planets orbiting younger stars than those in the original {\it Kepler} field.  {\it K2} observed several clusters across a wide range of ages, including the Upper Scorpius OB association (10My), Pleiades (115 My), and Hyades and Praesepe (600 My), compared to the stars in the {\it Kepler} field with a median age of $\sim$4 Gy. While there is tentative evidence from the small sample of planets detected so far that there is a measurable change in the occurrence rates of planets with time, as yet there is no robust confirmation of this trend. The NASA {\it TESS} mission will extend these observations further, observing young stars across the full sky. Importantly, we will soon have access to a statistically significant sample of planets orbiting stars whose ages span the timescales over which competing theories of planet formation and migration occur. \newline

\section{Planet radius evolution}
\label{sec:radius}
\vspace{-10pt}
Although the sample of planets transiting young stars is still small, we already see statistically significant differences from the planets in the {\it Kepler} field. Figure \ref{fig:rizzuto_k2} shows the radii of the seven detected planets orbiting young stars compared to the planets orbiting the older field population dwarfs in the same mass range. The radii are larger for the planets orbiting the younger stars, and larger still for the planet orbiting the youngest star. One possibility for the continuing inflation of planet radii even at 650~Myr is ongoing mass loss (Lopez et al. 2012). However, a statistically significant sample of planets with well-measured parameters and ages spanning stellar and planet characteristics is needed to clarify and understand the emerging trends. \newline

\begin{wrapfigure}{r}{0.45\linewidth}
\Centering
\includegraphics[width=0.45\textwidth]{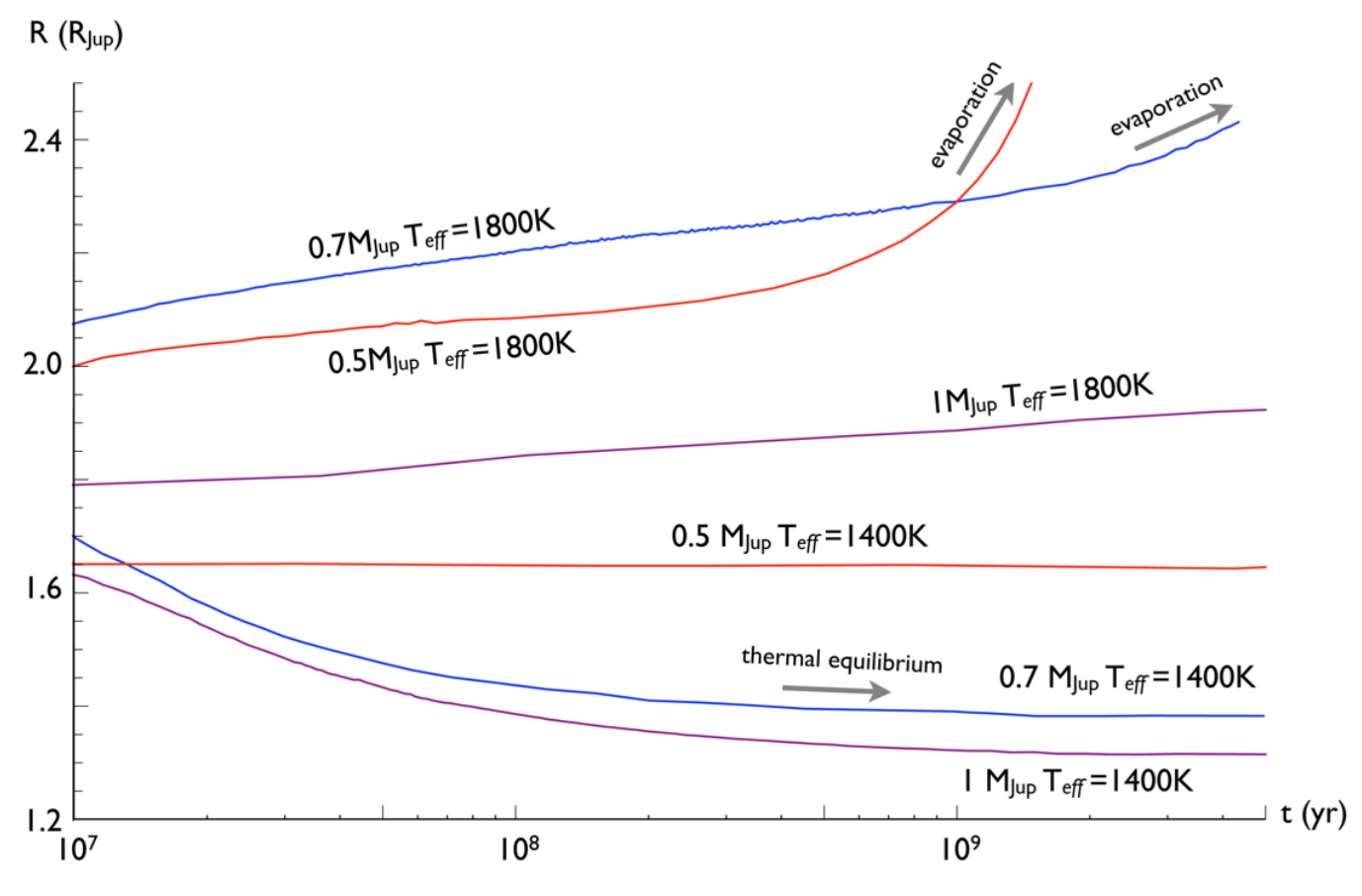}
\caption{Reproduced from Fig. 5 of Batygin et al. (2011), showing the radius evolution with time of several Ohmically heated hot Jupiter models.}
\label{fig:batygin2011}
\end{wrapfigure}

In addition to the inflated young planets, one conundrum that has emerged in the last two decades of exoplanet research is that of inflated hot Jupiters around field population stars. Going back to HD~209458b (the first hot Jupiter with a measured radius), these hot gas giants, well beyond the initial contraction phase, are larger than would be predicted from standard models for a given incident flux. There have been several suggestions for the mechanism by which these planets are inflated, many of which are reviewed in Fortney \& Nettelmann (2010). They include, but are not limited to, tidal heating due to ongoing orbital circularization (Bodenheimer et al. 2001; Leconte et al. 2010), conversion of stellar flux into kinetic energy (winds; Guillot \& Showman 2002), and Ohmic heating (Batygin \& Stevenson 2010; Batygin et al. 2011). Figure \ref{fig:batygin2011} shows an example of the time evolution of simulated hot Jupiters undergoing Ohmic heating. The models broadly fall into two categories, with planet radii either increasing with time or decreasing, depending on the initial mass and temperature. There is only a small fraction of initial conditions parameter space where the final equilibrium radius is equal to the initial radius, i.e. most of the planets are expected to change size with time if Ohmic heating is a dominant mechanism in governing the internal energy budget.

Planetary radius evolution is also a natural consequence of several other possible processes; e.g., tidal circularization injects additional energy into inflation of the atmosphere while the planet's orbit is circularizing, but once the eccentricity reaches zero the planet would deflate to its expected radius (Fortney \& Nettelmann 2010). One way to interrogate these theories would be to assemble a statistically significant sample of hot Jupiters with a set of precise ages, masses and radii. One could then compare the radii of those planets with the predictions of models (e.g., \ref{fig:batygin2011}) and observe whether the predicted trends are in fact present.

\vspace{10pt}As stars evolve off the main sequence, there is mounting observational evidence that planet radii begin to inflate again with the additional incident flux from the brightening star. Grunblatt et al. (2017) examine two gas giants orbiting evolved red giants stars, K2-97~b and K2-132~b, and show that their radii are consistent with the incident flux they are currently receiving, but inconsistent (inflated) with the flux they would have been receiving when the host star was on the main sequence. In addition, Grunblatt et al. (2018) analysed the three known planets transiting evolved stars and found they all have significantly non-zero eccentricity, well after the expected circularization timescale. They hypothesize that as the radius of the host star expands and the convection zone increases, the planets begin to tidally interact and inspiral into the star. As above, having a statistically significant sample of planets that span the interesting age range of main sequence star to sub-giant to red giant would allow us to pinpoint the driving physical processes that influence the evolution of the planets.

\section{Planet orbital evolution}
\label{sec:orbits}
\vspace{-10pt}
Another method to understand the possible evolutionary and migration processes experienced by planets is by observing the current spin-orbit alignments, or obliquities, of these systems (see also white paper by Marshall et al.). Migration by planet-planet scattering or by the Kozai mechanism can produce a range of obliquities which can circularize with time. However, disk migration processes are more likely to result in planets having more aligned orbits from early in the system lifetime. Based upon current obliquity measurements, stars cooler than $\sim$6100K are generally aligned with their orbiting gas giants, whereas stars hotter than $\sim$6100K display the full range of obliquities (Winn et al. 2010; Schlaufman et al. 2010). The break at this particular temperature is still unexplained, as early theories relating to the ability of the convective cores of the cooler stars to align the orbits of their planets through tidal dissipation would only apply to the close-in Jupiters. However, the trend holds for cooler Jupiters (Mazeh et al. 2015), which should be too far for tidal interactions to have a large impact.\newline

Recently, Yu et al. (2018) noted that one avenue of investigation is to focus on `dynamically young' Jupiters---those with periods from 5--10 days. These should experience weaker tidal effects, with circularization timescales as long as several Gyr, and therefore may still retain some of their primordial obliquity and eccentricity. By assembling a sample of well-characterized `dynamically young' Jupiters with accurate ages, it may be possible to map out the evolution of obliquities with time, and inform both the dominant migration processes and possibly address the question of the change in behaviour at $\sim$6100K. 

\section{Planet atmosphere evolution}
\label{sec:atmos}
\vspace{-10pt}
In addition to the evolution of the bulk properties of the planet population with time, the specific properties of the components of a given planet (atmosphere, surface, interior) are also expected to undergo dramatic shifts during a planetary lifetime. As planets further from their host stars radiate away the energy from their initial contraction, their temperature and luminosity decrease with time until they reach a temperature at equilibrium with the incident flux from the star. Even this process is poorly constrained---in our own solar system we are unable to explain Saturn's current luminosity, which is $\sim$50\% brighter than predicted (Fortney \& Nettelman 2010).\newline

\begin{wrapfigure}{r}{0.70\linewidth}
\Centering
\includegraphics[width=0.70\textwidth]{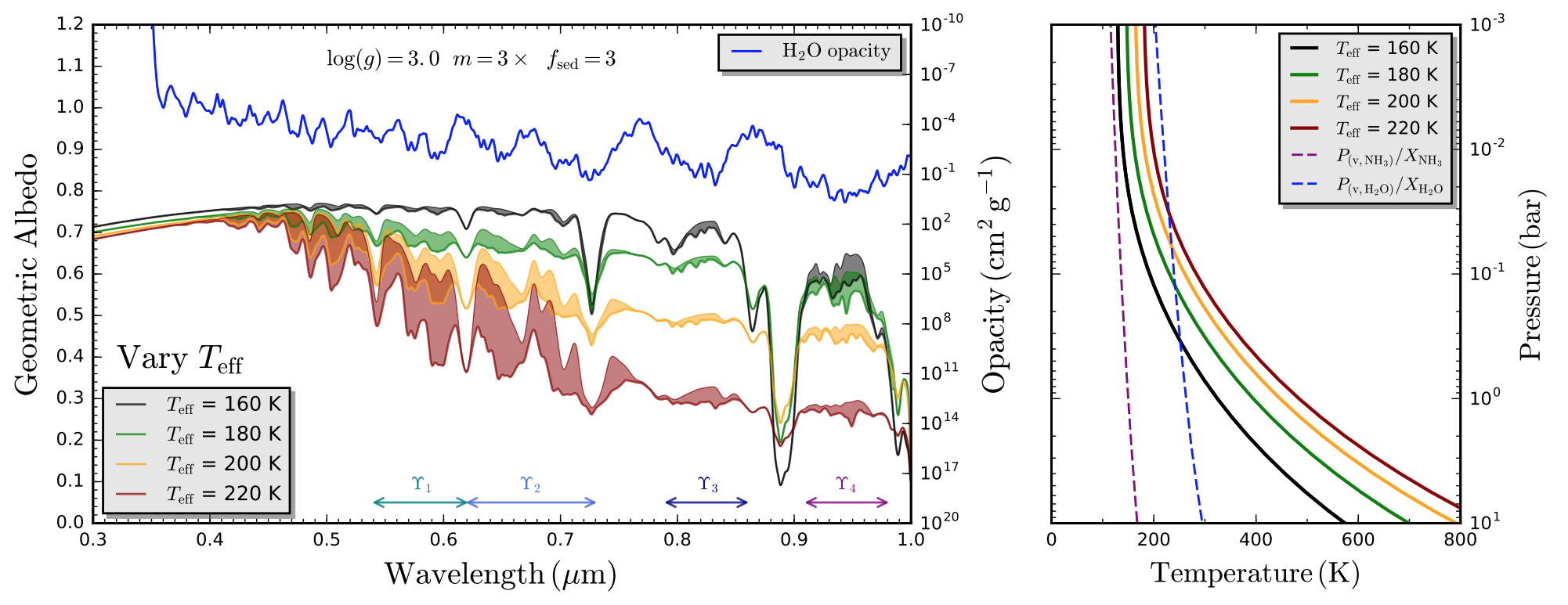}
\caption{Reproduced from Fig 5. of McDonald et al. (2018), showing the impact of changing temperature on reflected spectra from cool gas giants.}
\label{fig:mcdonald2018}
\end{wrapfigure}

As the planet cools, the atmosphere evolves and the observed properties (e.g., compositional abundance, atmospheric profiles) are expected to change. In particular, the planets will go through substantial cloud changes with time. The major contributors to these changes are CH$_4$ and H$_2$O. Figure \ref{fig:mcdonald2018} shows the impact of temperature on the modelled H$_2$O absorption features in the reflected light spectrum of cool gas giants. The planets are expected to become brighter (longward of 0.5$\mu m$) as they cool and reflective water clouds form out of the previously clear atmosphere. Cooler than 150~K, models predict that water clouds will sink deeper into the atmosphere and the troposphere will be dominated by NH$_3$ (ammonia) clouds (Sudarsky et al. 2000). Mapping out the atmospheric composition of cool gas giants as a function of time will be an important empirical test of these predictions.\newline

Finally, we note that the Earth's atmosphere has also undergone substantial changes with time, including the Great Oxygen Catastrophe at $\sim$2.4 Gyr. At most times in the past, the Earth would present a markedly different atmospheric spectrum than what would be observed today. Knowing the ages of the nearby bright stars which will be the targets of searches for Earth-like planets via direct imaging and spectroscopic measurements by future flagships will be of particular importance to understanding the frequency of Earth-like planets.


\section{Measuring stellar ages}
\label{sec:stars}
\vspace{-10pt}
All of the studies proposed above require the precise knowledge of the ages and parameters of the exoplanet host stars. The very high precision light curves produced by the NASA {\it Kepler} mission led to several breakthroughs in stellar physics via asteroseismology. The detection of oscillations in thousands of low-mass stars generated the unprecedented systematic measurement of fundamental stellar properties such masses, radii and ages with very high precision on a massive scale (Chaplin \& Miglio 2013). Of particular relevance, asteroseismic ages are now used to calibrate parameters derived from gyrochronology (van Saders et al. 2016) or magnetic activity (Metcalfe \& Egeland, 2019), and can probe stars as young as several hundred Myr (Lund et al. 2016). \newline

Obtaining well-calibrated asteroseismic parameters for a large number of exoplanet host stars will be crucial for understanding the formation and evolution of exoplanets. An accompanying white paper entitled ``Stellar Physics and Galactic Archaeology using Asteroseismology in the 2020's'' (Huber et al.), describes the prospects for the upcoming decade of asteroseismology. For exoplanets, this will include thousands of detections in dwarfs and subgiants by the NASA {\it TESS} mission, many of which will be exoplanet host stars. Asteroseismology with the Swiss-ESA \textit{CHEOPS} satellite will be able to determine ages for a few F and early G exoplanet hosts as well as for post-main sequence F,G and K stars (Moya et al 2018). In addition, \textit{Gaia} will help to assign TESS exoplanet host stars to clusters of known age even if TESS itself cannot determine an  asteroseismic age. Finally, ESA's {\it PLATO} mission will determine asteroseismic ages of nearly 100,000 dwarfs and subgiants a significant fraction of which will also host one or more exoplanets (Magrin et al 2018). Within the next decade, therefore,  we expect to assemble a large sample of exoplanets with well-defined ages with which to study planetary evolutionary processes.

\section{Conclusions}
\vspace{-10pt}
In this white paper we have presented a set of open questions in exoplanet formation, migration and evolution that are potentially accessible in the next ten years. We note that there are many time-dependent exoplanet processes that we have not covered here, particularly surface (e.g. ocean longevity, vulcanism) and sub-surface phenomena (e.g. tectonic activity, magnetic field changes), but these changes are not expected to be observable in the near-term. We suggest that a statistical sample of well-characterized exoplanet systems with precise ages will illuminate a wide variety of underlying formation and evolutionary processes. We are on the precipice of an influx of many thousands of new exoplanet discoveries from the NASA {\it TESS} mission, the ESA \textit{Gaia} and \textit{PLATO} missions, and the many ground-based transit, precision radial velocity and imaging surveys. Combined with the precise stellar parameters and ages for their host stars provided by the \textit{Gaia} and \textit{PLATO} missions respectively, we will have an unprecedented opportunity to make progress on these questions. 

\pagebreak
\noindent\textbf{References}

\noindent Batygin, K., Stevenson, D.~J., \& Bodenheimer, P.~H.\ 2011, ApJ, 738, 1\\
Bodenheimer, P., Lin, D.~N.~C., \& Mardling, R.~A.\ 2001, ApJ, 548, 466\\
Burrows, A., Hubbard, W.~B., Lunine, J.~I., \& Liebert, J.\ 2001, Reviews of Modern Physics, 73, 719\\
Chaplin, W.~J., \& Miglio, A.\ 2013, ARAA, 51, 353\\
Chiang, E., \& Laughlin, G. 2013, MNRAS, 431, 3444\\
Ford, E. B. 2006, PASP, 118, 364\\
Fortney, J.~J., \& Nettelmann, N.\ 2010, SSR, 152, 423\\
Goldreich, P., Lithwick, Y., \& Sari, R. 2004, ApJ, 614, 497\\
Guillot, T., \& Showman, A.~P.\ 2002, A\&A, 385, 156\\
Haisch, K. E., Jr., Lada, E. A., \& Lada, C. J. 2001, ApJL, 553, L153\\
Ida, S., \& Lin, D. N. C. 2008, ApJ, 673, 487-501\\
Ida, S., \& Lin, D. N. C. 2004, ApJ, 604, 388\\
Leconte, J., Chabrier, G., Baraffe, I., et al.\ 2010, A\&A, 516, A64\\
Lopez, E. D., Fortney, J. J., \& Miller, N. 2012, ApJ, 761, 59\\
Lund, M.~N., Basu, S., Silva Aguirre, V., et al.\ 2016, MNRAS, 463, 2600\\
Magrin, D., Ragazzoni, R., Rauer, H., et al.\ 2018, Space Telescopes and Instrumentation 2018: Optical, Infrared, and Millimeter Wave, 10698, 106984X \\
Mazeh, T., Perets, H. B., McQuillan, A., \& Goldstein, E. S. 2015, ApJ, 801, 3\\
Metcalfe, T. S., \& Egeland, R. 2019, ApJ, 871, 39\\
Moya, A., Barcel{\'o} Forteza, S., Bonfanti, A., et al.\ 2018, A\&A, 620, A203\\ 
Pollack, J. B., Hubickyj, O., Bodenheimer, P., et al. 1996, Icarus, 124, 62\\
Rizzuto, A.~C., Vanderburg, A., Mann, A.~W., et al.\ 2018, AJ, 156, 195\\
Schlaufman, K. C., Lin, D. N. C., \& Ida, S. 2009, ApJ, 691, 1322\\
Schlaufman, K. C. 2010, ApJ, 719, 602\\
Sudarsky, D., Burrows, A., \& Pinto, P.\ 2000, ApJ, 538, 885\\
van Saders, J. L., Ceillier, T., Metcalfe, T. S., et al. 2016, Nature, 529, 181\\
Winn, J. N., Fabrycky, D., Albrecht, S., \& Johnson, J. A. 2010, ApJ, 718, L145\\
Yu, L., Zhou, G., Rodriguez, J.~E., et al.\ 2018, AJ, 156, 250\\

\end{document}